\documentstyle[11pt,newpaspcno,epsf]{article}

\newcommand{\CI}{C\,{\sc i}}
\newcommand{\NI}{N\,{\sc i}}
\newcommand{\OI}{O\,{\sc i}}
\newcommand{\SI}{S\,{\sc i}}
\newcommand{\FeI}{Fe\,{\sc i}}
\newcommand{\NiI}{Ni\,{\sc i}}
\newcommand{\FeII}{Fe\,{\sc ii}}
\newcommand{\water}{\mbox{\rm H$_2$O}}
\newcommand{\feh} {\mbox{\rm [Fe/H]}}
\newcommand{\ch} {\mbox{\rm [C/H]}}
\newcommand{\oh} {\mbox{\rm [O/H]}}

\newcommand{\nfe} {\mbox{\rm [N/Fe]}}
\newcommand{\ofe} {\mbox{\rm [O/Fe]}}
\newcommand{\nc} {\mbox{\rm [N/C]}}
\newcommand{\co} {\mbox{\rm [C/O]}}

\newcommand{\forCI} {\mbox{\rm [C\,{\sc i}]}}
\newcommand{\forOI} {\mbox{\rm [O\,{\sc i}]}}
\newcommand{\teff}  {\mbox{$T_{\rm eff}$}}

\newcommand{\logg}  {\mbox{{\rm log}\,$g$}}

\begin{document}
\title{CNO at the surface of low and intermediate MS stars\altaffilmark{1}}

\author{Poul E. Nissen}
\affil{Department of Physics and Astronomy, University of Aarhus, Denmark} 




\altaffiltext{1}{Partly based on observations collected at the European Southern
Observatory, Chile (ESO No. 67.D-0106)}
\setcounter{footnote}{1}


\begin{abstract}

Recent studies of spectral lines of the CNO elements in the Sun and unevolved
disk and halo stars show that most published abundance data may be affected
by systematic errors due to inadequate 1D modelling of stellar
atmospheres. This raises doubts about previously derived trends of CNO
abundances as a function of \feh . In this review we concentrate on
abundance ratios derived from weak spectral features with similar dependences
on temperature and pressure, which ratios are then relatively insensitive to
3D model atmosphere effects and to \teff\ and surface gravity.
The recent controversy about the trend of \ofe\ vs. \feh\ is discussed,
new results for C/O ratios as determined from forbidden lines and high 
excitation neutral lines are presented, and N/C ratios derived from NH and
CH lines are commented on. 

\end{abstract}




%
%
%

\section{Introduction}
In this review I will discuss CNO abundances of main-sequence (MS) stars
with masses between 0.5 and 2 solar masses. Such stars have convective 
envelopes,
which means that any pollution of the stellar atmosphere caused by 
accretion of interstellar gas will be diluted by mixing in the envelope. 
The convective envelope is, on the other hand, not deep enough to bring up
products from nuclear reactions to the surface of the star, except in the
case of Li and Be, which elements may be destroyed by proton
reactions at the bottom of the convective zone in the less massive stars.
Hence, it seems rather safe to assume, that CNO at the surface of low and 
intermediate MS stars can be used as a tracer of CNO abundances
in the interstellar gas and dust out of which the stars were formed. By deriving
abundances of MS stars with different ages and belonging to different
populations we may then obtain important information about the nucleosynthesis
and Galactic evolution of the CNO elements.

The assumption of undisturbed chemical composition in the atmospheres of
cooler MS stars should, however, be carefully checked. As shown by
Lebreton et al. (1999), a comparison of CM diagrams based on Hipparcos
parallaxes of nearby, unevolved stars with stellar model computations,
suggests a mild depletion of heavier elements due to microscopic diffusion
at the bottom of the convective envelope. According to the 
computations of Morel \& Baglin (1999), the effect is of the order of
0.10 dex for metal-poor stars with ages of 10 Gyr. It is likely,
however, that the corresponding effect on abundance ratios such as C/O and
N/C is considerably smaller and may be neglected in comparison with
other possible systematic errors arizing from analysis of spectral lines.

In the following, we will first take a look at the recent controversy about the
trend of \ofe\ vs. \feh . Then follows some new results on the \co\ - \oh\
relation compared to previous results and a discussion of the \nc\ - \feh\
relation stressing the need for new high resolution studies of nitrogen
abundances in cooler MS stars. In particular, I will emphasize the large
granulation effects on the temperature and pressure structure of 
metal-poor stellar atmospheres, which have to be taken into account when 
deriving abundances from spectral lines or molecular bands.
 
\section{Oxygen}
According to standard nucleosynthesis models, oxygen is exclusively made by
supernovae of Type II, whereas iron is also produced by Type Ia SN when
they start to occur after a time delay of 0.5 - 1 Gyr. Hence, the O/Fe
ratio can be used as a chemical clock to date the star formation process.
If $\ofe \simeq +0.4$ the major star formation episodes must mave occurred
within a time interval of $\la 0.5$\,Gyr; if $\ofe \simeq 0.0$ the 
time interval of star formation has been $\ga 1$\,Gyr. Par example, standard
models of the chemical evolution of the Galaxy predicts a near constant
$\ofe \simeq +0.4$ for metallicities below $\feh \simeq -1.0$, perhaps with
an increase of \ofe\ below $\feh = -2.0$ due to a higher O/Fe yield ratio
in the most massive supernovae (Chiappini et al. 1999).

Extensive studies of the forbidden oxygen lines in K giants have
resulted in a near-constant $\ofe \simeq +0.4$ for halo stars 
with $-2.5 < \feh < -1.0$ (e.g. Barbuy 1988, Kraft et al. 1992).
More recently, Sneden \& Primas (2001) have confirmed this; \ofe\
is not higher that +0.5 dex at $\feh = -3.0$. Israelian et al. (1998, 2001)
and Boesgaard et al. (1999), on the other hand, find a linear rize of
\ofe\ with decreasing \feh\  when oxygen abundances are derived from
the near-UV OH lines in spectra of MS stars using 1D model atmospheres.
At $\feh = -3.0$ \ofe\ reaches values of 1.0 to 1.2
dex. The view has been expressed that the \forOI -based
results for the giants are in error on two accounts:
(i) oxygen is reduced in the atmospheres of
giants because the convective envelope has mixed oxygen-poor
material to the surface from the ON-cycled interior, and (ii)
the \forOI\ lines are partially filled in by chromospheric emission.
These explanations seem, however, unlikely: (i) the strong nitrogen enhancement
resulting from the ON cycle has not been seen, and (ii) no asymmetries of the
\forOI\ lines have been detected.

\begin{figure}
\plotfiddle{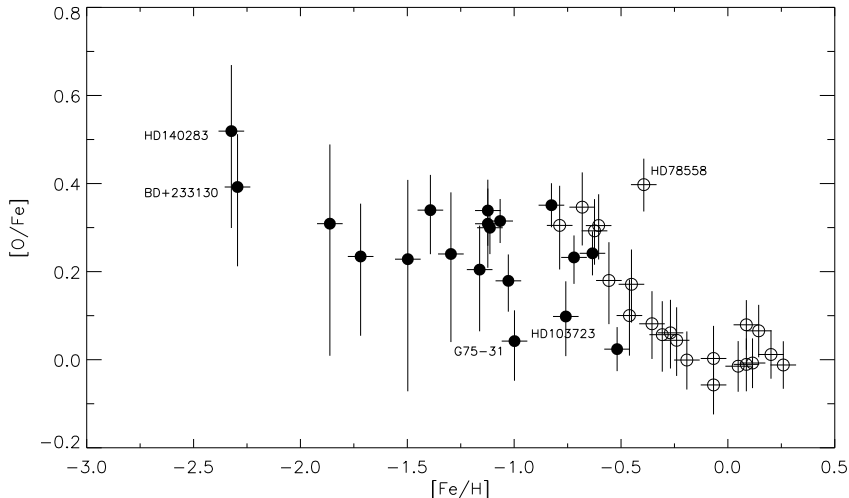}{7.0cm}{0}{90}{90}{-170}{0}
\caption{\small{\ofe\ vs. \feh\ as derived from the \forOI\,6300 \AA\ line
and \FeII\ lines in spectra of MS stars and two subgiants, HD\,140283 and
BD\,+23\,\,3130. Approximate correction for the effects
of stellar granulation has been applied.
Filled circles refer to halo stars from Nissen et al. (2002) and open circles
to disk stars from Nissen \& Edvardsson (1992)}}
\end{figure}

In order to advance on this oxygen problem, Nissen et al. (2002) have recently
studied the faint \forOI\,6300 \AA\ line in metal-poor MS star spectra
obtained with the VLT/UVES instrument. The spectra have resolutions of
60\,000 to 120\,000 and a very high S/N of 300 - 600. Equivalent widths as
small as 0.5\,m\AA\ could be measured with an error of $\pm 0.2$\,m\AA .
Iron abundances were derived from a number of weak \FeII\ lines. The data
were first analyzed with 1D MARCS models (Asplund et al. 1997) 
using \teff\ values
determined from $b-y$ and $V-K$, and gravities derived via Hipparcos
parallaxes and/or
the Str\"{o}mgren $c_1$ index. An important point is that the derived
\ofe\ ratio is only weakly dependent on possible errors in \teff\ and
practically independent of errors in the gravity. 

In the analysis of the \forOI\,6300 \AA\ line, Nissen et al. included the
effect of a blending \NiI\ line, which recently was shown by Allende 
Prieto, Lambert
\& Asplund (2001) to decrease the derived solar oxygen abundance by 0.13\,dex.
As this \NiI\ line has nearly disappeared when $\feh < -1.5$ , the
net effect is to increase the derived \ofe\ by about 0.1 dex for 
metal-poor halo stars. Furthermore, a study of the influence of stellar 
granulation was undertaken by applying the new generation of 3D, 
time-dependent hydrodynamical model atmospheres (Asplund et al. 1999, 2000).
Due to the expansion of rising granulation elements and the lack of radiative
heating in metal-poor stellar atmospheres they have much lower temperature
and electron pressure in the upper layers than classical 1D models in 
radiative equilibrium. The \forOI\ 6300\,\AA\ line is formed high up in the
atmosphere and are stronger in 3D models than in 1D, whereas the \FeII\
lines are formed deeper and are computed to be slightly weaker in 3D.
The net 3D effect is a decrease of \ofe\ ranging
from 0.1\,dex at $\feh = -1.0$ to about 0.2\,dex at $\feh = -2.5$.

Fig.\,1 shows the derived \ofe\ values after having applied the 3D corrections.
The figure also includes disk stars from Nissen \& Edvardsson (1992)
with their \ofe\ values corrected for the presence of the \NiI\ blend.
As seen there is a rather sharp rize of \ofe\ around $\feh \simeq -0.5$
to a constant level of $\ofe \simeq 0.3$ in the range $-2.0 < \feh < -0.7$.
The two subgiants, HD\,140283 and BD\,+23\,\,3130, give a hint of an 
increase of \ofe\ below $\feh = -2.0$
but this has to be confirmed by more data. Hence, the new results
for MS and subgiant stars essentially agree with previous results from
the forbidden oxygen lines in giant star spectra.

As shown by Asplund \& Garc\'{\i}a P\'{e}rez (2001) the 3D effects on 
oxygen abundances derived from near-UV OH lines are as large as 
$-0.6$\,dex for turnoff stars with $\feh \simeq -3.0$ due to the fact
that molecule lines are  very sensitive to the temperature structure of
the stellar atmosphere. This explains the high \ofe\ values derived by
Israelian et al. (1998, 2001) and Boesgaard et al. (1999) on the basis of
a classical 1D model atmosphere analysis. All problems
are, however, not solved. As discussed by Nissen et al. (2002) oxygen abundances
in MS stars derived from the \OI\ 7774\,\AA\ triplet tend to be higher than
abundances from the forbidden line even when non-LTE and 3D corrections
are applied. It also remains to be seen if oxygen abundances
derived from the OH IR lines (Balachandran, Carr \& Carney 2001; 
Mel\'{e}ndez, Barbuy, \& Spite 2001; Mel\'{e}ndez \& Barbuy 2002)
and the forbidden lines in giants will agree with the trend shown in
Fig.\,1 when 3D effects are taken into account.

A further problem is that the trend of \ofe\ with \feh\ may not be
universal. In Fig.\,1 there are several stars in the halo-disk transition
region ($-1.0 < \feh < -0.5$) that deviate 2 - 3 sigma from the mean trend.
Furthermore, Nissen \& Schuster (1997) have clearly shown that a group of
six halo stars with $-1.0 < \feh < -0.7$ have lower 
$\alpha$-element/Fe ratios (including O/Fe) than thick disk stars and
other metal-poor halo stars in the same metallicity range. The six halo
stars move in large Galactic orbits and might have been
accreted from dwarf galaxies or have been formed in low-density regions
in the outer halo with a chemical evolution that has proceeded more slowly
than in the inner halo and the thick disk (Matteucci \& Fran\c{c}ois 1992).

\section{Carbon}
The most accurate carbon abundances of lower MS stars are obtained from the
weak forbidden \CI\ line at 8727\,\AA . Recently, Allende Prieto,
Lambert \& Asplund
(2002) determined the solar C abundance from this line and obtained 
log$\epsilon$(C) = 8.39 $\pm 0.04$\,dex taking into account 3D model
atmosphere effects and a revised $gf$ value of the line. In combination
with their solar oxygen abundance log$\epsilon$(O) = 8.69 $\pm 0.05$\,dex
(Allende Prieto et al. 2001) the solar atmospheric C and O abundances now agree
well with C and O abundances in early-type MS stars and in the local ISM.

\begin{figure}
\plotfiddle{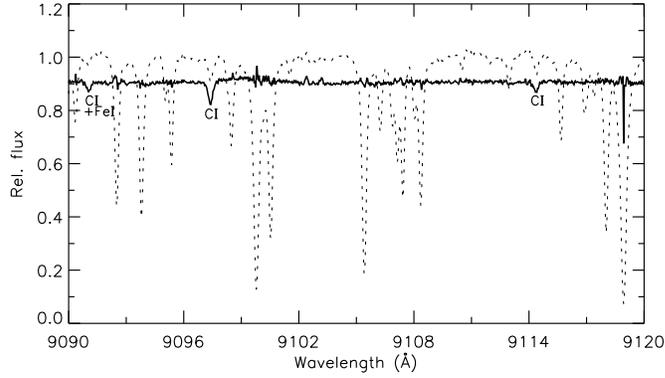}{5.0cm}{0}{100}{100}{-135}{0}
\caption{\small{The VLT/UVES spectrum of CD$-$35\,14849 (\teff\ = 6120\,K,
\logg\ = 4.1 and $\feh = -2.3$). The dotted line shows
the spectrum before removal of the numerous telluric \water\ lines. The thick,
full drawn line is the spectrum after division with the spectrum of
the B-type star HR\,8858 using the IRAF task `$\em telluric$' to obtain the
best fit between the two sets of \water\ lines. 
Excessive noise in the stellar spectrum are seen
where strong \water\ lines have been removed. The weak \CI\ lines are
marked; one of them being blended by a \FeI\ line}}
\end{figure}

\begin{figure}
\plottwo{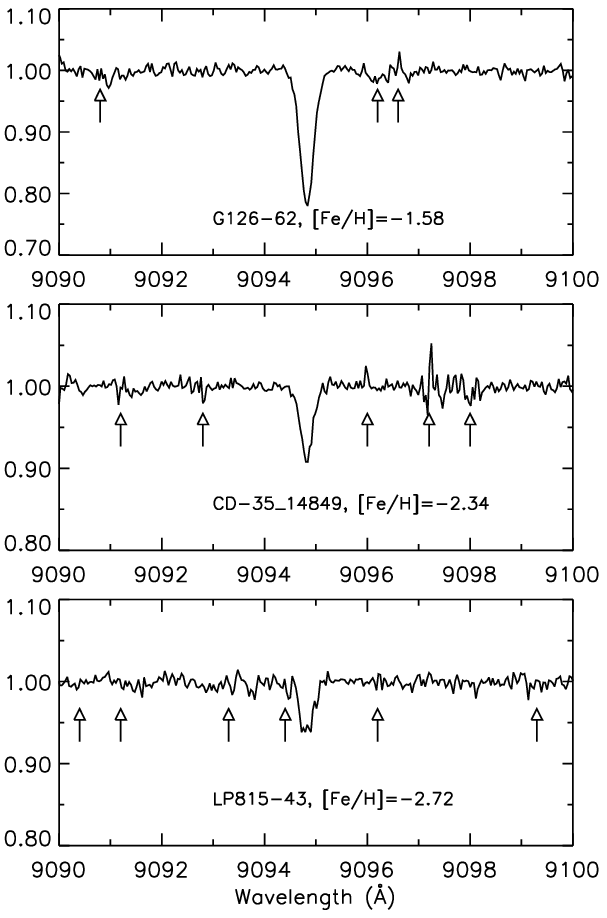}{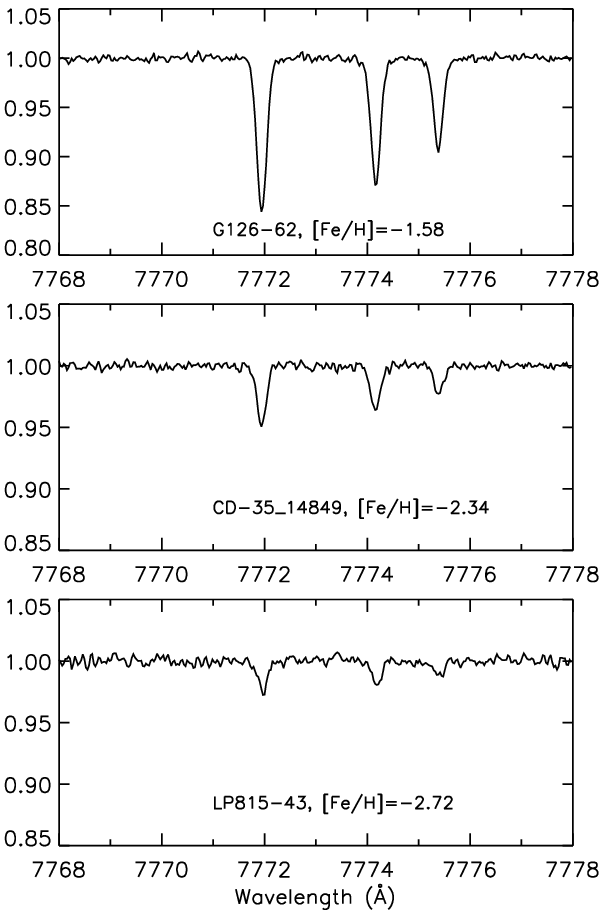}
\caption{\small{A sequence of VLT/UVES spectra of metal-poor turnoff
stars around the
$\lambda$9094.9 \CI\ line (left) and the \OI\ triplet (right).
The arrows indicate positions where telluric
\water\ lines have been removed}}
\end{figure}

The \forCI\ 8727\,\AA\ line was observed in disk stars by Anderson \& Edvardsson
(1994) and Gustafsson et al. (1999). [C/Fe] as derived from 1D models
shows an increase of about 0.2\,dex when going from solar metallicity 
to $\feh \simeq -1$. In view of the differential 3D model atmosphere
corrections 
to be expected some of this encrease may, however, be spurious. It is
probably more safe to compare the derived C abundance with the O abundance
derived from the \forOI\ 6300\,\AA\ line. Both lines arise from low excitation
levels and are formed in the same layers of stellar atmospheres. Hence,
the derived C/O ratio is insensitive to errors in \teff\ and surface gravity,
and the 3D corrections for C and O are expected to be
nearly the same. This is confirmed for the Sun; Allende Prieto et al. 
(2001, 2002)
find a 3D correction of $-0.08$ dex for both elements. I have therefore
combined 
the carbon abundances from Anderson \& Edvardsson (1994) and Gustafsson 
et al. (1999) with oxygen abundances determined from the \forOI\ $\lambda 6300$
line as measured by Nissen \& Edvardsson (1992) including corrections for
the \NiI\ blend discussed in Sect. 2. The derived \co\ values are shown in Fig.\,4. 

The \forCI\ 8727\,\AA\ line is too weak to be measured in halo MS stars
with $\feh < -1$. Instead one may use four high excitation \CI\ lines around
9100\,\AA\ and the \OI\ triplet at 7774\,\AA\ to
derive C/O. This technique was first applied by Tomkin et al. (1992), who
correctly emphasized that the derived abundance ratio is insensitive
to errors in stellar atmospheric parameters and the $T,P$ structure of the
model atmosphere, i.e. 3D hydrodynamical effects. Furthermore, non-LTE
corrections are relatively small and go in the same direction
for the \CI\ and
\OI\ lines. As seen from their Fig.\,10, \co\ is nearly constant at the
level of $-0.5$\,dex in the range $-2.5 < \feh < -1.0$.

In order to test and expand the C/O results of Tomkin et al. (1992),
I have derived C and O abundances from near-IR VLT/UVES spectra
originally obtained together with M. Asplund and M. Pettini to determine
sulphur abundances from the \SI\ 9212 - 9237\,\AA\ triplet.
The spectra have $R = 60\,000$ and $S/N \sim$ 200
to 300. The equivalent widths of the \OI\ triplet can be very
accurately determined (see Fig.\,3), whereas the \CI\ lines around
9100\,\AA\ fall in a spectral
region with many strong telluric \water\ lines. As seen from Fig.\,2, these
lines may, however, be removed by dividing with a spectrum of a hot star,
and Fig.\,3 shows that the \CI\ line at 9094\,\AA\ can be clearly observed
down to metallicities of about $-3$ in halo turnoff stars.

\begin{figure}
\plotfiddle{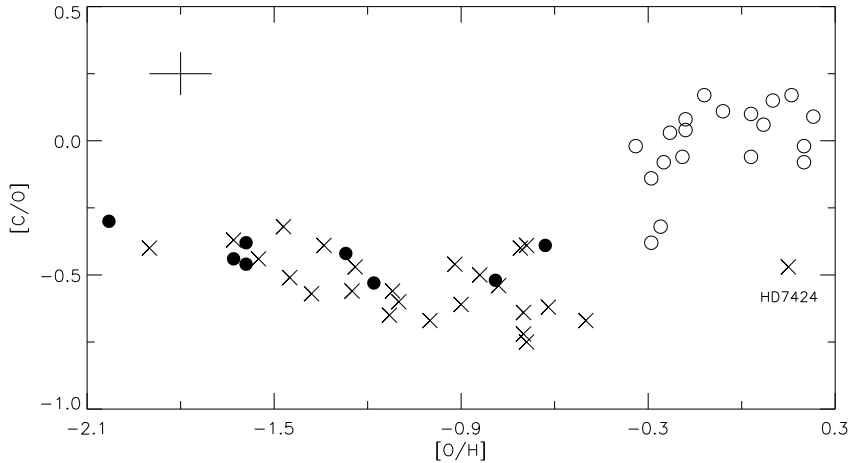}{7.0cm}{0}{90}{90}{-170}{0}
\caption{\small{\co\ vs. \oh . Open circles are disk stars with C and O
abundances derived from forbidden lines; see text for references. 
Filled circles (VLT/UVES data) and crosses (Tomkin et al. 1992) are halo
stars with C and O abundances derived from permitted high excitation
\CI\ and \OI\ lines. Typical 1-sigma error bars, $\pm 0.08$ on \co\ and 
$\pm 0.10$ on \oh , are shown in the upper left corner}}
\end{figure}

The \co\ ratios derived from a 1D, LTE model atmosphere analysis of
the UVES spectra are shown
in Fig.\,4 together with data from Tomkin et al. (1992) and the disk star
results discussed above. The disk star analysis of the forbidden C and O
lines was made differentially with respect to the Sun; hence
the adopted solar abundances do not matter. In the analysis of the UVES 
spectra of the high excitation permitted \CI\ and \OI\ lines 
the new solar C and O abundances of Allende Prieto et al. (2001, 2002),
log$\epsilon$(C) = 8.39 and log$\epsilon$(O) = 8.69, were adopted. 
Tomkin et al. (1992), on the other hand, used the older values
of Grevesse et al. (1991), log$\epsilon$(C) = 8.60 and log$\epsilon$(O) = 8.92.
The solar C/O ratio is, however, practically the same, so no offset in
\co\ is present. None of the data have been corrected for non-LTE
or 3D effects, because these effects are expected to cancel in the
C/O ratio as explained above. As seen from Fig.\,4 \co\ is nearly constant
at the level of $-0.5$\,dex
among the halo stars perhaps with a slight decrease when going from
$\oh = -2.0$ to $\oh = -0.5$. Around $\oh = -0.3$, \co\ is rising steeply
to the solar abundance ratio. It will be interesting to see how well
Galactic chemical evolution models can explain
this upturn of \co\ in terms of metal-enhanced 
production of carbon in massive stars (Maeder 1992) and/or 
delayed production of carbon in intermediate mass stars.
In this connection, one should note the peculiar position of HD\,7424
in Fig.\,4, an oxygen rich but carbon poor star with halo kinematics,
which gives a first hint of a dichotomy in the evolution of \co\ between
the Galactic halo and the disk.

\section{Nitrogen}
Nitrogen abundances of intermediate and low main-sequence stars are
more difficult to determine than C and O abundances. The forbidden
N lines are very weak and have not been detected even in the solar spectrum.
Weak high excitation \NI\ lines can be measured in spectra of disk
stars. The studies of Clegg, Lambert \& Tomkin (1981) and Shi, Zhao
\& Chen (2002) suggest that N follows Fe, i.e. $\nfe \sim 0.0$ in the
metallicity range
$-0.8 < \feh < +0.1$, but the dispersion is quite large, which may be caused 
by errors in the observations of the weak lines. 

In the case of halo stars the high excitation \NI\ lines cannot be
detected and molecular lines of NH or CN have to be applied for 
the determination of N abundances. In order to be independent of the carbon
abundance determination the best choice is the NH band at
3360\,\AA . Bessell \& Norris (1982) were the first to make a high-resolution
survey and spectrum synthesis analysis of this feature. Among 25
halo dwarfs they discovered two stars (HD\,74000 and HD\,160617) with very
high N abundances, i.e. $\nfe \sim 1.7$ - 2.0. Tomkin \& Lambert (1984)
made a careful high resolution study of the $\lambda 3360$ NH band
relative to the $\lambda 4300$ CH band in spectra of 14 halo and disk 
stars with $-2.3 < \feh < -0.3$. They noted that the CH and NH hydrides
have almost identical dissociation energies; hence the circumstances of
CH and NH line
formation are similar and the derived N/C abundance ratio is insensitive
to atmospheric parameters and structure. As an important result they
found $\nc \simeq 0.0$ and concluded that nitrogen behaves like a
primary element with respect to carbon down to a metallicity of
$\feh \sim -2$. This conclusion is, however, based on only five halo stars.

More extensive studies of N abundances in lower MS stars
have been carried out by Laird (1985) and Carbon et al. (1987),
in both cases based on intermediate resolution spectra of
the NH $\lambda$\,3360 band and a 1D model atmosphere synthesis.
The conclusion is that N tend to follow Fe,
i.e. $\nfe \sim constant$ except for a small decline of \nfe\ for
\feh\ below $-1.5$ in the work of Carbon et al. The two studies 
are, however, puzzling in some respects. Laird's data have an unexplained
offset $\Delta \nfe \sim -0.6$ with respect to Carbon et al., 
and in both studies there is a clear dependence of the derived \nfe\
on \teff . Furthermore, one can expect large downward 3D corrections
of the N abundances derived from the NH band like in the case of oxygen
abundances derived from the OH lines.

\section{Conclusions}
Progress has been made in understanding recent discrepancies
on the \ofe\ - \feh\ relation. It looks like \ofe\ values derived 
from a 1D model atmosphere analysis of the near-UV OH lines are much too
high (Asplund \& Garc\'{\i}a P\'{e}rez 2001).
From a 3D analysis of the \forOI\ $\lambda$\,6300 line in MS stars
\ofe\ seems to be near constant at a
level of $\ofe \simeq 0.3$ for halo stars with $-2 < \feh < -1$
(Nissen et al. 2002). It remains, 
however, to be seen if oxygen abundances derived for subgiant and
giant stars from the \OI\ triplet and the IR OH lines will be in agreement,
and the trend of \ofe\ below $\feh \sim -2$ has to be
studied in detail.

New results for the C/O ratio show an interesting trend of \co\ vs. \oh\
with a strong upturn of \co\ at $\oh \simeq -0.3$. More data should be 
obtained especially for stars in the halo-disk transition region to
look for a possible dichotomy in the evolution of C/O in the Galactic halo and
the disk.

Nitrogen appears to have the least reliable abundances of the CNO elements.
Existing data for \nfe\ are affected by unexplained systematic errors,
and large downward 3D corrections of N abundances from the 
NH $\lambda$\,3360 band are to be expected for stars at low metallicities.
Hence, we 
very much need a new high resolution study of the NH $\lambda$\,3360 lines
in Galactic halo stars coupled with a careful 3D model atmosphere study.
The recent discovery of large variations of N/O ratios among oxygen poor
damped Lyman $\alpha$ systems (e.g. Pettini et al. 2002) accentuates the need 
for more reliable data
and hence a better understanding of the nucleosynthesis of nitrogen.


\begin{references}
\reference 
Allende Prieto, C., Lambert, D.L., \& Asplund, M. 2001, ApJ, 556, L63
\reference 
Allende Prieto, C., Lambert, D.L., \& Asplund, M. 2002, ApJ, 573, L137
\reference 
Anderson, H., \& Edvardsson, B. 1994, A\&A, 290, 590
\reference 
Asplund M., Gustafsson B., Kiselman D., \& Eriksson K. 1997, A\&A, 318, 521
\reference 
Asplund M., Nordlund \AA ., Trampedach R., \& Stein R.F. 1999, A\&A, 346, L17
\reference 
Asplund, M., Ludwig, H.-G., Nordlund, \AA., \& Stein, R.F. 2000, A\&A, 359, 669
\reference 
Asplund M., \& Garc\'{\i}a P\'{e}rez A.E. 2001, A\&A, 372, 601
\reference
Balachandran, S.C., Carr, J.S., \& Carney, B.W. 2001, New Astronomy Rev., 45,
529
\reference 
Barbuy, B. 1988, A\&A, 191, 121
\reference 
Bessell, M.S., \& Norris, J. 1982, ApJ, 263, L29
\reference 
Boesgaard, A.M., King, J.R., Deliyannis, C.P., \& Vogt, S.S. 1999, AJ 117, 492
\reference 
Carbon, D.F., Barbuy, B., Kraft, R.P. et al. 1987, PASP, 99, 335
\reference 
Chiappini, C., Matteucci, F., Beers, T.C., \& Nomoto, K. 1999, ApJ, 515, 226
\reference 
Clegg, R.E.S., Lambert, D.L., \& Tomkin, J. 1981, ApJ, 250, 262
\reference 
Grevesse, N., Lambert, D.L., Sauval et al. 1991, A\&A, 242, 488
\reference 
Gustafsson, B., Karlsson, T., Olsson, E., Edvardsson, B., \& Ryde, N. 1999,
A\&A, 342, 426
\reference 
Israelian, G., Garc\'{\i}a Lopez, R.J., \& Rebolo, R. 1998, ApJ, 507, 805
\reference 
Israelian G., Rebolo, R., Garc\'{\i}a Lopez R.J., et al. 2001 , ApJ, 551, 833
\reference 
Kraft, R.P., Sneden, C., Langer, G.E., \& Prosser, C.F. 1992, AJ, 104, 64
\reference 
Laird, J.B. 1987, ApJ, 289, 556
\reference 
Lebreton, Y., Perrin, M.-N., Cayrel, R., Baglin, A., \& Fernandes, J.
1999, A\&A, 350, 587
\reference 
Maeder, A. 1992, A\&A, 264, 105
\reference 
Matteucci, F., \& Fran\c{c}ois, P. 1992, A\&A, 262, L1
\reference 
Mel\'{e}ndez, J., \& Barbuy, B. 2002, ApJ, 575, 474
\reference 
Mel\'{e}ndez, J., Barbuy, B., \& Spite, M. 2001, ApJ, 556, 858
\reference 
Morel, P., \& Baglin, A. 1999, A\&A, 345, 156
\reference 
Nissen, P.E., \& Edvardsson, B. 1992, A\&A, 261, 255
\reference 
Nissen P.E., \& Schuster W.J. 1997, A\&A, 326, 751
\reference
Nissen P.E., Primas F., Asplund M., \& Lambert D.L. 2002, A\&A, 390, 235
\reference 
Pettini, M., Ellison, S.L., Bergeron, J., \& Petitjean, P. 2002, A\&A 391, 2002
\reference 
Shi, J.R., Zhao, G., \& Chen, Y.Q. 2002, A\&A, 381, 982
\reference 
Sneden, C., \& Primas, F. 2001, New Astronomy Rev., 45, 513
\reference 
Tomkin, J., \& Lambert, D.L. 1984, ApJ, 279, 220
\reference 
Tomkin, J., Lemke, M., Lambert, D.L., \& Sneden, C. 1992, AJ, 104, 1568
\end{references}
\end{document}